\documentclass[10pt]{article} 

\usepackage{amsmath} % amssymb package, useful for mathematical
\usepackage{amssymb}

\usepackage[pdftex]{graphicx}

\usepackage{cite}

\usepackage{color}

\usepackage[labelfont=bf,labelsep=period,justification=raggedright]{caption}

\makeatletter
\renewcommand{\@biblabel}[1]{\quad#1.}  
\makeatother

\begin{document}

% Title must be 150 characters or less
\begin{flushleft} {\Large \textbf{Composite structural motifs of
binding sites for delineating biological functions of proteins} }\\

% Insert Author names, affiliations and corresponding author email.  \\
Akira R. Kinjo$^{1\ast}$, Haruki Nakamura$^{1}$, \\ \textbf{1}
Institute for Protein Research, Osaka University, Suita, Osaka
565-0871, Japan.  \\ $\ast$ E-mail: akinjo@protein.osaka-u.ac.jp \\
Short title: Composite motifs of binding sites
\end{flushleft}

% Please keep the abstract between 250 and 300 words
\section*{Abstract} Most biological processes are described as a
series of interactions between proteins and other molecules, and
interactions are in turn described in terms of atomic structures. To
annotate protein functions as sets of interaction states at atomic
resolution, and thereby to better understand the relation between
protein interactions and biological functions, we conducted exhaustive
all-against-all atomic structure comparisons of all known binding
sites for ligands including small molecules, proteins and nucleic
acids, and identified recurring elementary motifs.  By integrating the
elementary motifs associated with each subunit, we defined composite
motifs which represent context-dependent combinations of elementary
motifs. It is demonstrated that function similarity can be better
inferred from composite motif similarity compared to the similarity of
protein sequences or of individual binding sites. By integrating the
composite motifs associated with each protein function, we define
meta-composite motifs each of which is regarded as a time-independent
diagrammatic representation of a biological process. It is shown that
meta-composite motifs provide richer annotations of biological
processes than sequence clusters.  The present results serve as a
basis for bridging atomic structures to higher-order biological
phenomena by classification and integration of binding site
structures.

% Please keep the Author Summary between 150 and 200 words 
% Use first person. PLoS ONE authors please skip this step.  
% Author Summary not valid for PLoS ONE submissions.

\section*{Introduction} Virtually every biological process is
realized, at the atomic level, through a series of interactions
between proteins and other molecules.  Accordingly, most proteins, if
not all, synchronously or asynchronously interact with multiple
molecules ranging from single atom ions, small (non-polymer) molecules
to proteins, nucleic acids and other macromolecules. The types and
combinations of interactions in proteins are known to modulate their
functions. For example, depending on their ligand-bound or ligand-free
forms as well as interactions with corepressor or coactivator
proteins, nuclear receptors can perform intricate transcriptional
regulations \cite{SantosETAL2011}.  The activity of coronavirus
3C-like proteases is controlled by dimerization through their
C-terminal domain which is absent from picornavirus 3C proteases
\cite{ShiETAL2004}. Furthermore, certain homologous proteins can
catalyze completely different enzymatic reactions, namely transferase
or hydrolase activities, by adopting different oligomerization states
\cite{KoikeETAL2009}.  Therefore, it is important to identify not only
individual interactions, but also possible combinations of the
interactions that can be accommodated by a protein to fully describe
its molecular functions as well as to distinguish different functions
among homologous proteins.

The advance in genome sequence technologies is making it more and more
imperative to develop effective techniques for inferring protein
functions from sequence information. To date, the most widely used
approach for protein function prediction is the annotation transfer,
which is based on the assumption that protein functions are similar if
their sequences are similar
\cite{Friedberg2006,LoewensteinETAL2009,RentzschANDOrengo2009}. It has
been gradually recognized, however, that such annotation transfer
approaches may be unreliable in many cases
\cite{Rost2002,SchnoesETAL2009}. It has also been shown that function
similarity is not a simple function of sequence similarity
\cite{LouieETAL2009}.  These observations prompt us to have more
detailed examination of the determinants of protein functions.

Structural information has been proved valuable for precisely
understanding protein functions \cite{BrandenANDTooze1999}.  Thanks to
the structural genomics efforts \cite{Levitt2007,TerwilligerETAL2009},
we now have a great wealth of structural information available for
close examination of sequence-structure-function relationships of
proteins. However, when global topologies or folds of protein
structures are considered, it is often even more difficult to assign a
specific function to a particular fold, for some folds include an
extremely diverse set of proteins with diverse functions
\cite{KoikeETAL2009, NaganoETAL2002}.  The use of structural
information is not limited to finding global fold similarity and
distant evolutionary relationship.  In particular, physical
interactions between protein molecules and their ligands observed in
experimentally solved protein structures allow more direct approaches
to elucidate the relationship between protein structures and functions
\cite{KinjoANDNakamura2009,KinjoANDNakamura2010}. To date, there have
been many methods for detecting potential ligand binding sites based
on structural similarity of proteins
\cite{ChenETAL2011,WallaceETAL1997,StarkETAL2003,KinoshitaANDNakamura2003,GoldANDJackson2006,BrylinskiANDSkolnick2008,XieANDBourne2008,KinjoANDNakamura2009}. Most
of these methods are targeted at predicting protein functions at the
level of ligand binding and catalytic activity. There have also been
many studies on protein-protein interaction interfaces to understand
biological functions of proteins in cellular contexts
\cite{KeskinETAL2004,JaninETAL2008,KeskinETAL2008,SCOPPI,PPiClust,TyagiETAL2009,KinjoANDNakamura2010,ZhaoETAL2011,RobinsonETAL2007,DasguptaETAL2011,TuncbagETAL2008,SinhaETAL2010,TuncbagETAL2011}.
However, apart from a few works \cite{DavisANDSali2010,Davis2011,RausellETAL2010}, most of these studies are focused on particular types of
interactions \emph{per se} and do not explicitly address how 
the combination of interactions with small molecules and macromolecules 
modulates with biological function of proteins. 

To understand the relationship between the patterns of interactions at
atomic level and biological functions of proteins, we herein performed
exhaustive all-against-all structural comparisons of binding site
structures at atomic level using all structures available in the
Protein Data Bank (PDB) \cite{wwPDB}, and identified recurring
structural patterns of ligand binding sites to define \emph{elementary
motifs}. We then defined \emph{composite motifs} by integrating the
elementary motifs associated with individual subunits. In other words,
protein subunits with the same combination of elementary motifs are
said to share an identical composite motif.  We then examined how such
composite motifs correlated with protein functions as defined by the
UniProt database \cite{UniProt}.  It is demonstrated that the
similarity between composite motifs better corresponds with the
similarity between functions compared to the similarity between
protein sequences or between individual binding sites.  Finally, by
integrating all the composite motifs associated with particular
functions, we define \emph{meta-composite motifs}. It is shown that
meta-composite motifs are useful to elucidate the rich internal
structures of biological processes compared to sets of homologous
sequence clusters.

% Results and Discussion can be combined.
\section*{Results}
\label{sec:results}
\subsection*{Identification of elementary and composite motifs}
\label{comp}

We first generated all biological units as annotated in the PDBML
\cite{PDBML} files, and then extracted 197,690 protein subunits which
contained at least one ligand (non-polymer, protein or nucleic acid)
binding site.  Here, a ligand binding site of a subunit is defined as
a set of atoms of the subunit that are in contact with some atoms of
the ligand within 5\AA{}.  While we do not use any pre-defined
non-redundant data set based on sequence similarity, the redundancy is
taken care of after clustering similar structures (see below). In this
manner, the structural diversity of proteins with highly homologous or
identical amino acid sequences can be preserved in the following
analyses while the structural redundancy is removed.

\begin{figure}[!ht]
  \begin{center}
  \includegraphics[width=.8\textwidth]{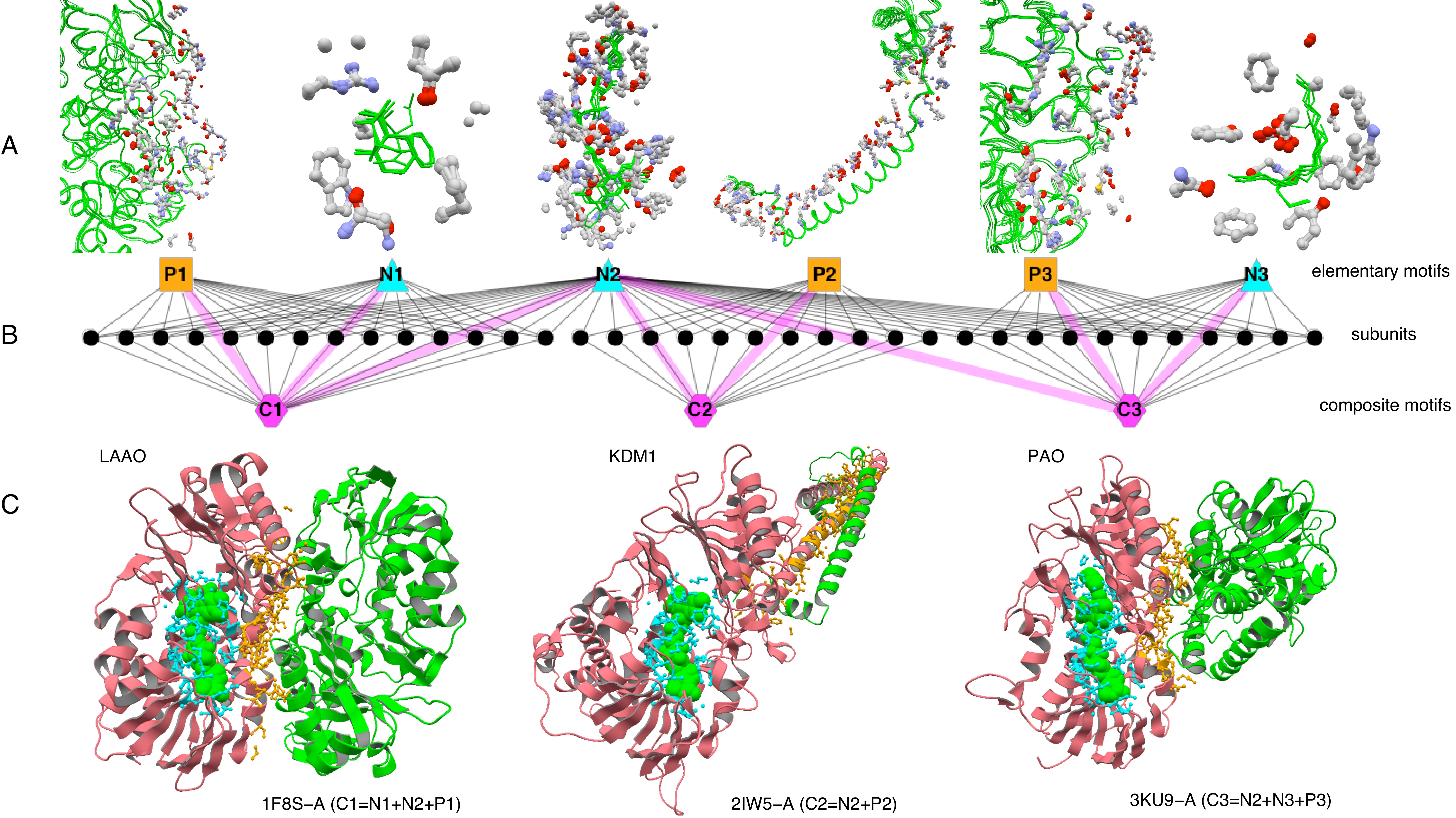}    
  \end{center}
  \caption{\label{fig:motif} \textbf{Examples of elementary and composite motifs.}
A: Concrete examples of elementary motifs (corresponding to
B). Several binding sites belonging to each elementary
motif are superimposed. The binding site atoms that constitute the
elementary motif are shown in ball-and-stick representation with CPK
coloring and ligands are shown in green wireframes (non-polymers) or
tubes (proteins). These binding sites include subunits shown in C.
Non-polymer ligands are phenylalanine and its analogs (N1), FAD (N2), and 
polyamines (N3).
B: In this example, the combinations of 3 non-polymer binding
elementary motifs (cyan triangles labeled N1, N2 and N3) and 3 protein
binding elementary motif (orange rectangles labeled P1, P2 and P3)
found in various protein subunits (black dots) define 3 distinct
composite motifs (hexagons in magenta labeled C1, C2, and C3). 
Examples of each elementary motif are shown in molecular figures (A)
right above the triangles or rectangles, and those of each
composite motif are shown in molecular figures (C) right below the
hexagons.
Direct correspondence between elementary and composite motifs is indicated by
thick edges in pale magenta.
C: Concrete examples of composite motifs (corresponding to B). 
These 3 composite motifs share the same elementary motif for FAD binding 
(labeled N2 in B).
Subunits (colored pink) containing the composite motifs (C1, C2, C3)
are shown with elementary motifs in ball-and-stick representations
(protein binding sites in orange, non-polymer binding sites in cyan) and
with ligands in green (spacefill for non-polymers, cartoon for
proteins).
From left to right: 
L-amino acid oxidase (LAAO) from \emph{Calloselasma rhodostoma} in homo-dimeric form (PDB ID: 1F8S \cite{1F8S}, chain A); 
human lysine-specific histone demethylase 1 (KDM1) (PDB ID: 2IW5 \cite{2IW5}, chain A);
polyamine oxidase (PAO) from \emph{Zea mays} in putative homo-dimeric form (PDB ID: 3KU9 \cite{3KU9}, chain A, pdbx\_struct\_assembly.id 3).
The protein figures were created using jV \cite{KinoshitaANDNakamura2004}.
The network diagrams (also in Figs. \ref{fig:metacomp} and \ref{fig:eg}) were created using Cytoscape \cite{Cytoscape}.
}
\end{figure}

All-against-all structure comparisons of 410,254 non-polymer binding
sites, 346,288 protein binding sites and 20,338 nucleic acid binding
sites using the GIRAF structure search and alignment program
\cite{KinjoANDNakamura2007} followed by complete linkage clustering
yielded 5,869, 7,678 and 398 clusters (with at least 10 members) of
non-polymer, protein and nucleic acid binding sites, respectively.
(We did not use in the following analyses small clusters with less
than 10 members because some small clusters exhibited spurious
similarities.)  We refer to these clusters as \emph{elementary motifs}
in the following. An elementary motif can be regarded as a bundle of
mutually similar atomic dispositions of binding sites
(Fig. \ref{fig:motif}A). It should be noted that the elementary motifs
are solely based on the binding site structures, and they do not
directly include the identity of the binding partners.  We have
previously performed comprehensive analyses of elementary motifs
\cite{KinjoANDNakamura2009,KinjoANDNakamura2010}. It was found that
most elementary motifs were confined within homologous families. In
some exceptional cases, motifs were shared across non-homologous
families with different folds, which included motifs for metal,
mononucleotide or dinucleotide binding for non-polymer binding sites
\cite{KinjoANDNakamura2009} and coiled-coil motifs for protein binding
sites \cite{KinjoANDNakamura2010}.

The set of all elementary motifs contained in a protein subunit is
called the \emph{composite motif} of the subunit
(Fig. \ref{fig:motif}B,C). Thus, two subunits sharing the same set of
elementary motifs are said to have the same composite motif. In total,
5,738 composite motifs, each of which is shared by at least 10
subunits, were identified. Our hypothesis is that thus defined
composite motifs exhibit good correspondence with protein
functions. In the example in Fig \ref{fig:motif}, while the three
proteins (LAAO \cite{1F8S}, KDM1 \cite{2IW5} and PAO \cite{3KU9})
share the same elementary motif (N2) for FAD binding and they share
the same domain folds (FAD/NAD(P)-binding domain and FAD-linked
reductases C-terminal domain \cite{SCOP}), their biological functions
are similar but different; and these differences correspond to
the differences in their composite motifs.

\subsection*{Characterization of composite motifs} 　
\begin{figure}[!ht]
  \begin{center}
  \includegraphics[width=.4\textwidth]{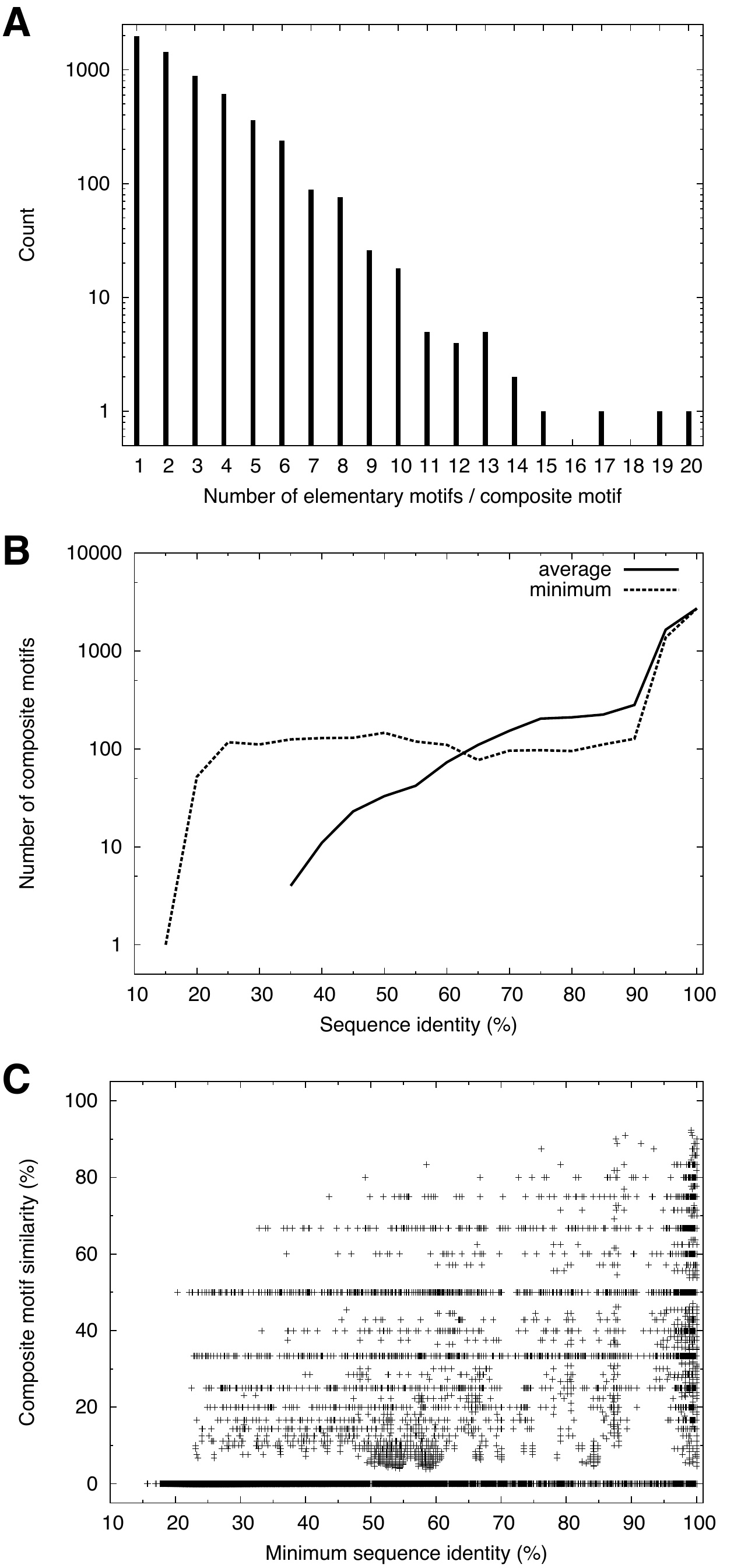}    
  \end{center}

  \caption{\label{fig:chara} \textbf{Characterization of composite motifs.}
A: Histogram of the number of elementary motifs comprising composite motifs. 
B: Histograms of the average and minimum sequence identities (\%) between
pairs of subunits within each composite motif. 
C: Composite motif similarity as a function of minimum sequence
identity between pairs of composite motifs. Sequence identity between
two composite motifs is defined as the sequence identity between two
protein sequences, one belonging to the one motif, the other to the
other motif.
}
\end{figure}

The number of elementary motifs that comprise a composite motif ranges
from 1 to 20 (Fig. \ref{fig:chara}A). Approximately one third of the
composite motifs (1975 out of 5738) consist of only one elementary
motif and more than 90\% of the composite motifs are composed of less
than or equal to 5 elementary motifs. The number of composite motifs appears 
exponentially decreasing as the number of constitutive elementary motifs 
increases.

To characterize the diversity of composite motifs, the average and
minimum sequence identities were calculated for pairs of subunits
sharing the same composite motifs (Fig. \ref{fig:chara}B).  Although
the majority of composite motifs are shared between close homologs on
average, many of them contain distantly related subunits. In
particular, 118 composite motifs were shared between subunits whose
sequence similarity could not be detected by BLAST
\cite{AltschulETAL1997}.  However, only three out of these 118
composite motifs consisted of more than one and at most two elementary
motifs.  Thus, if a composite motif consists of more than one
elementary motif, it is most likely to comprise only homologous
proteins.

By defining the similarity between two composite motifs as the
fraction of shared elementary motifs (Eq. \ref{eq:jac}), 
we also examined the similarity
between different composite motifs as a function of minimum sequence
identity between them (Fig. \ref{fig:chara}C). While many composite
motifs share no elementary motifs for the entire range of sequence
identities, some do share a significant fraction of their constitutive
elementary motifs in spite of weak sequence similarities. It is also
noted that the composite motif similarities widely vary for high
sequence identities. Thus, while each composite motif comprises
homologous proteins in most cases, the converse does not hold
in general so that composite motif similarity hardly correlates with
sequence similarity. This observation clearly demonstrates that  
it is not possible to take into account the structural diversity of 
binding sites and their combinations by using a representative set of 
proteins based on sequence similarity.

\subsection*{Association of composite motif similarity with  function similarity}

\begin{figure}[!ht]
  \begin{center}
  \includegraphics[width=.8\textwidth]{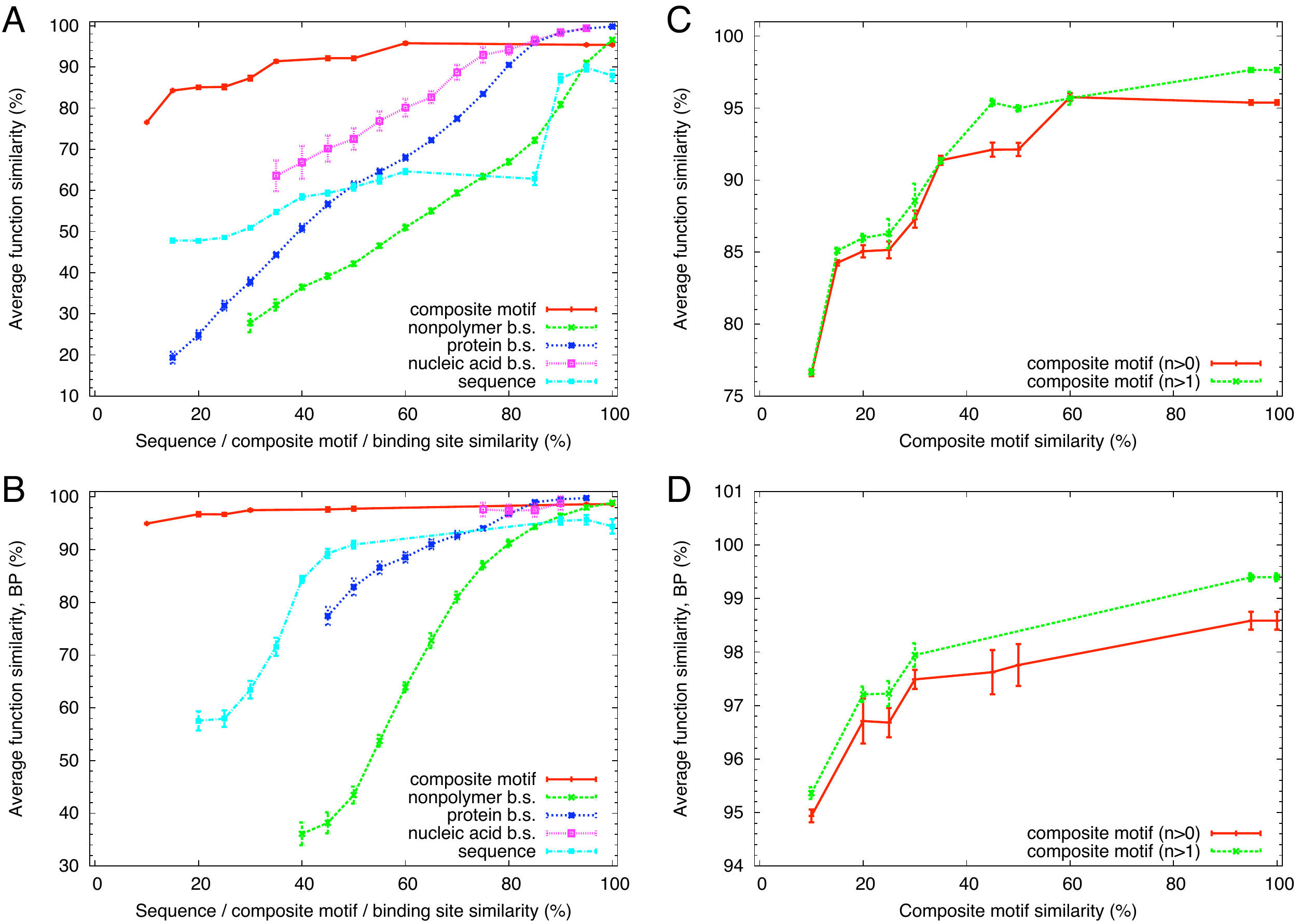}
  \end{center}

  \caption{\label{fig:pred} \textbf{Correspondence between composite motifs and protein functions.}
A: Average UniProt function similarity as a function of similarity
between subunits based on composite motifs, individual binding sites or
sequence identity.  Data points with insufficient number of samples
were discarded (see Materials and Methods). Error bars indicate the standard deviation of the average 
function similarity based on 10 bootstrap samplings.
B: Same as A, except that only the UniProt functions
of the Biological process category were used.
C: Composite motifs with more than one elementary motif (n$>$1) are compared with those with at least one elementary motif (n$>$0), the latter are the same as in A.
D: Same as C, except that only the UniProt functions of the Biological process category were used.}
\end{figure}

In order to study the functional relevance of the composite motifs, we
next examined the association between composite motif similarity and
function similarity. Here, the function of a protein is defined as a
set of controlled keywords provided in UniProt \cite{UniProt} and the
similarities for composite motifs and UniProt functions are defined by
the Jaccard index (see Materials and Methods, Eq. \ref{eq:jac}).  
For comparison, we
also checked sequence identity as well as binding site similarity 
(Eq. \ref{eq:simif}) as measures of subunit similarities in place of 
composite motif
similarity (Fig. \ref{fig:pred}A). In order to reduce the bias due to 
the redundant data set, we randomly
picked one representative from each composite or elementary motif, or
sequence cluster (with 100\% sequence identity cutoff) for this comparison. 
It is evident
that the function similarity persists even for low composite motif
similarities although the function similarity is not always 100\% for
100\% composite motif similarity. To the contrary, we can only
infer high function similarities for high sequence or binding site
similarities.

Since many UniProt function annotations, especially
those of ligand binding activities, have been actually derived from
the PDB entries, the high correlation between composite motifs and
UniProt functions may appear trivial. However, the current elementary
motifs that constitute composite motifs do not directly include the
information of their ligands, but are solely based on their binding
site structures. The bare binding site similarity does not correspond
with the function similarity as strongly as the composite motif
similarity.  In addition, when we used only the UniProt functions
under the Biological process category which are less directly related
to molecular functions, we still observed the highest function
similarity for a wide range of composite motif similarity compared to
sequence or binding site similarities (Fig. \ref{fig:pred}B).  These
results demonstrate that composite motifs sharing a small fraction of
elementary motifs imply more function similarity compared to bare
sequence or binding site similarities.

When we examined the correspondence between composite motifs and
UniProt functions excluding those composite motifs that consisted of
only one elementary motif, the correspondence was found to be slightly
better (Figs. \ref{fig:pred}C,D). This indicates that combinations of multiple
elementary motifs may enhance accurate inference of specific protein
functions.

Although the similarity between composite motifs implies similar
functions, 15 composite motifs were found to be shared by completely
different functions.  11, 3, and 1 of these composite motifs consisted
of 1, 2, and 3 elementary motifs, respectively. 7 of them were due to
improper annotations for artificially engineered proteins, to
incomplete annotations in the UniProt, or to a wrong annotation in the
PDB, and 3 were due to coiled-coil structures. Among the remaining 5
composite motifs, 2 composite motifs were actually found in the same
dimeric complexes, and each of them consisted of only 1 elementary
motif shared between remotely homologous proteins.

\subsection*{Examples of composite motifs sharing the same elementary 
motif and fold but with different functions}
\begin{figure}[!ht]
  \begin{center}
  \includegraphics[width=.8\textwidth]{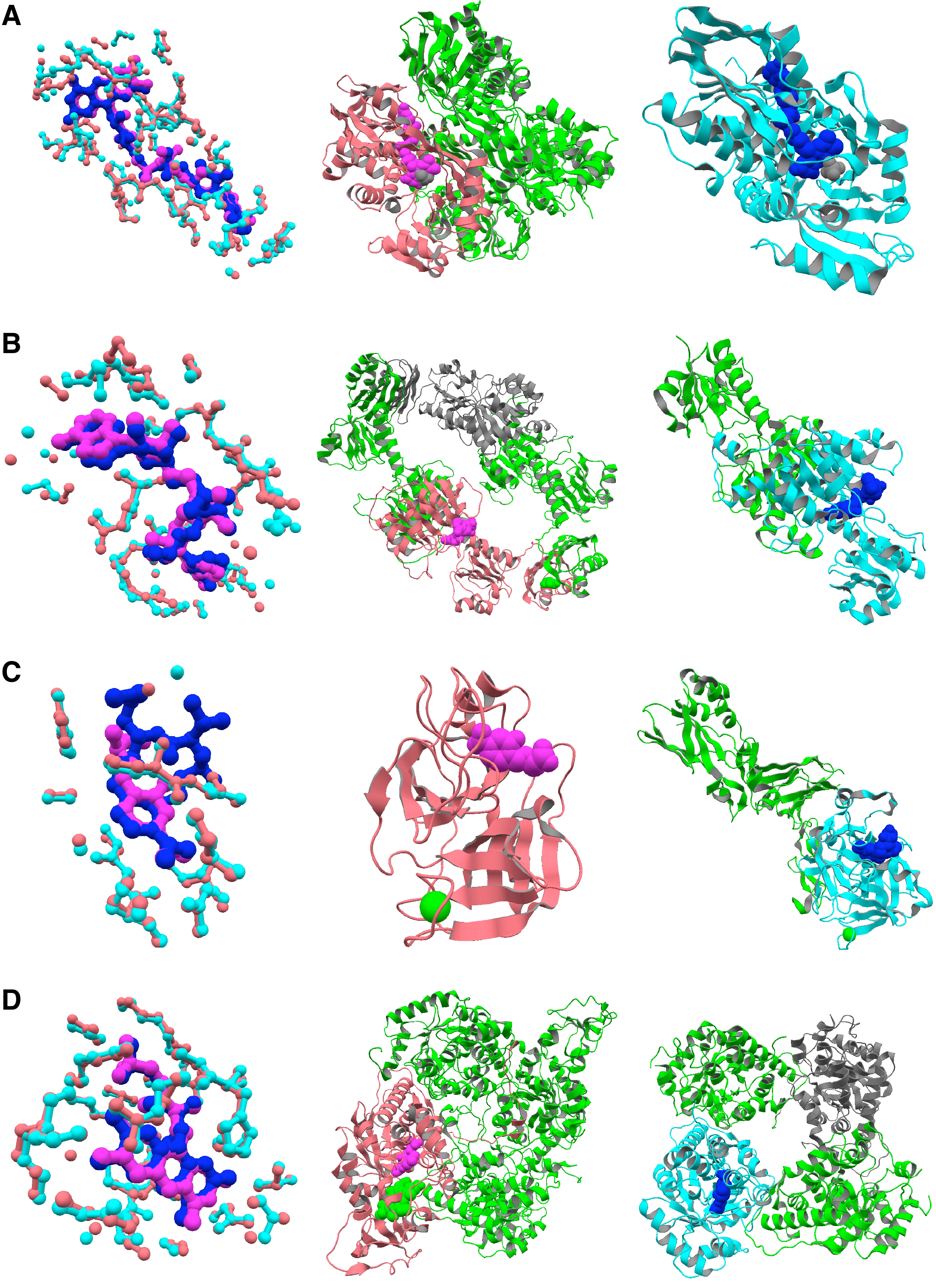}    
  \caption{\label{fig:examples} 
\textbf{Examples of differences in composite motifs and functions.}
Left column: superposition of common elementary motifs (pink and cyan) and their ligands (magenta and blue).
Center column: the biological unit containing the subunit with the elementary motif shown in the left column in pink, with interacting molecules (other than that in the left column) in green and non-interacting molecules in grey.
Right column: the biological unit containing the subunit with the elementary motif shown in the left column in cyan, with interacting molecules (other than that in the left column) in green and non-interacting molecules in grey.
A: Glycine oxidase (center) and glycerol-3-phosphate dehydrogenase (right), sharing FAD binding motif (left).
B: D-3-phosphoglycerate dehydrogenase (center) and C-terminal binding protein 3 (right) sharing NAD binding motif (left).
C: $\beta$-trypsin (center) and coagulation factor VII (right) sharing protease inhibitor binding motif (left).
D: Cytochrome $b_2$ (center) and glycolate oxidase (right) sharing FMN binding motif (left).
}
  \end{center}
\end{figure}

We have already presented in Introduction an example that demonstrated
different combinations of elementary motifs (i.e., composite motifs)
might modulate function specificity (Fig. \ref{fig:motif}).  The
analysis in the previous section showed that composite motif
similarity is a good indicator of function similarity. In this
section, we provide several examples of proteins that share the same
elementary motif and the same fold, but have different composite
motifs and different functions (Fig. \ref{fig:examples}). 
These examples show that the
difference in functions can be associated with the difference in
composite motifs within the same family of proteins.

\paragraph{Glycine oxidase (GO) and glycerol-3-phosphate dehydrogenase (GlpD)}
% N0:33 FAD
% 1RYI1   2984041 A-1     13      1955    {N0:33,N1547:0,P0:1168,P1704:0,P1705:0} GLYCINE OXIDASE
% 2QCU1   4588921 A-1     14      4874    {N0:33,N3265:0} Aerobic glycerol-3-phosphate dehydrogenase

GO from \emph{Bacillus subtilis} (PDB 1RYI \cite{1RYI}, chain A) and GlpD from
\emph{Escherichia coli} (PDB 2QCU \cite{2QCU}, chain A) share the same 
elementary
motif for binding the FAD cofactor, and despite the low sequence
similarity ($\sim$ 14\% sequence identity), they share the same fold
(FAD/NAD(P)-binding domain \cite{SCOP}) according to the Matras fold
comparison program \cite{KawabataANDNishikawa2000,Kawabata2003} (Fig. \ref{fig:examples}A).
While GO forms a homotetramer and has 3 elementary motifs for protein
binding, GlpD is monomeric (however, the latter may form a
dimer \cite{2QCU}). In addition, they have their own elementary motif
for binding the respective ligands (glycolic acid, GOA, in PDB 1RYI and
phosphoenolpyruvate, phosphate ion, PO4, in PDB 2R46). Thus, they have different
composite motifs.  Although the shared elementary motif for FAD
binding and the shared fold, they exhibit different enzymatic
activities, EC 1.4.3.19 for GO and EC 1.1.5.3 for GlpD, and function in
different contexts, thiamine biosynthesis and glycerol
metabolism, respectively.

\paragraph{D-3-phosphoglycerate dehydrogenase (PGDH) and C-terminal-binding protein 3 (CtBP3)}
% N0:27 NAD
% 1PSD1   2882773 A-1     9       1950    {N0:27,N228:1,P0:842,P0:851}    D-3-PHOSPHOGLYCERATE DEHYDROGENASE (PHOSPHOGLYCERATE DEHYDROGENASE)
% 1HKU1   2499769 A-1     3       2606    {N0:27,P0:1642} C-TERMINAL BINDING PROTEIN 3

PGDH from \emph{E. coli} (PDB 1PSD \cite{1PSD}, chain A, EC 1.1.1.95)
and CtBP3 (also called CtBP1) from rat (PDB 1HKU \cite{1HKU}, chain A,
EC 1.1.1.-) share the same elementary motif for binding the NAD
cofactor and the same folds (NAD(P)-binding Rossmann-fold domain and
Flavodoxin-like fold \cite{SCOP}) with 25 \% sequence identity (Fig. \ref{fig:examples}B).  PGDH
forms a homotetramer with one of its protein-protein interface located
at its additional ACT domain \cite{ACTdomain}, and is involved in
L-serine biosynthesis.  CtBP3, forming a homodimer or heterodimer with
CtBP2, is involved in controlling the structure of the Golgi complex
and acts as a corepressor targeting various transcription 
regulators \cite{1HKU}. While these proteins may catalyze very similar 
reactions, their biological roles are clearly different.

\paragraph{$\beta$-trypsin and coagulation factor VII}
% N11:4 inhibitor
% 1G3C1   2430433 A-1     3       1825    {N11:4,N25:0}   BETA-TRYPSIN
% 1WQV1   3207421 B-1     18      4534    {N11:4,N25:8,P0:556,P1330:0}    Coagulation factor VII
Bovine $\beta$-trypsin (PDB 1G3C \cite{1G3C}, chain A, EC 3.4.21.4)
and human coagulation factor VII heavy chain (PDB 1WQV \cite{1WQV},
chain H, EC 3.4.21.21) are both serine proteases with 40 \% sequence
identity. In these structures, they share the same elementary motif for 
protease inhibitors at the catalytic sites in addition to similar calcium 
ion binding sites
although the latter do not belong to the same elementary motif (Fig. \ref{fig:examples}C).
Factor VII heavy chain is in complex with its light chain counter
part as well as with tissue factor, which shapes its functional form. 
On the other hand, $\beta$-trypsin is not known to form 
a similar complex structure. Thus, the difference in their complex structures 
can be associated with the difference in their functions: 
digestion for $\beta$-trypsin and blood coagulation for Factor VII.

\paragraph{Cytochrome $b_2$  and glycolate oxidase (GOX)}
% N7:1 FMN
% 1FCB1   2395405 B-1     17      523     {N7:1,P0:514,P0:550,P159:0}     FLAVOCYTOCHROME B2
% 1AL71   2173645 A-1     9       3814    {N7:1,P0:1640,P159:0}   GLYCOLATE OXIDASE
Mitochondrial cytochrome $b_2$, also known as L-lactate dehydrogenase, 
from \emph{Saccharomyces cerevisiae}
(PDB 1FCB \cite{1FCB}, chain A, EC 1.1.2.3) and glycolate oxidase
(GOX) from spinach (PDB 1AL7 \cite{1AL7}, chain A, EC 1.1.3.15) share the
TIM-barrel fold with 40 \% sequence identity, and have the same
elementary motif for flavin mononucleotide (FMN)
(Fig. \ref{fig:examples}D).  Although they have roughly equivalent
homotetrameric complexes, the number of interacting subunits are
different: a subunit of cytochrome $b_2$ interacts with all 3 other
subunits whereas that of GOX interacts with only 2 out of 3 other
subunits.  In addition, cytochrome $b_2$ also has an elementary motif
for heme binding in its additional heme-binding domain which is utilized for 
transferring electrons to cytochrome $c$ following oxidation of lactate \cite{1FCB};
such function is not associated with GOX.

\subsection*{Meta-composite motifs for annotating functions} 
\begin{figure}[!ht]
  \begin{center}
  \includegraphics[width=.8\textwidth]{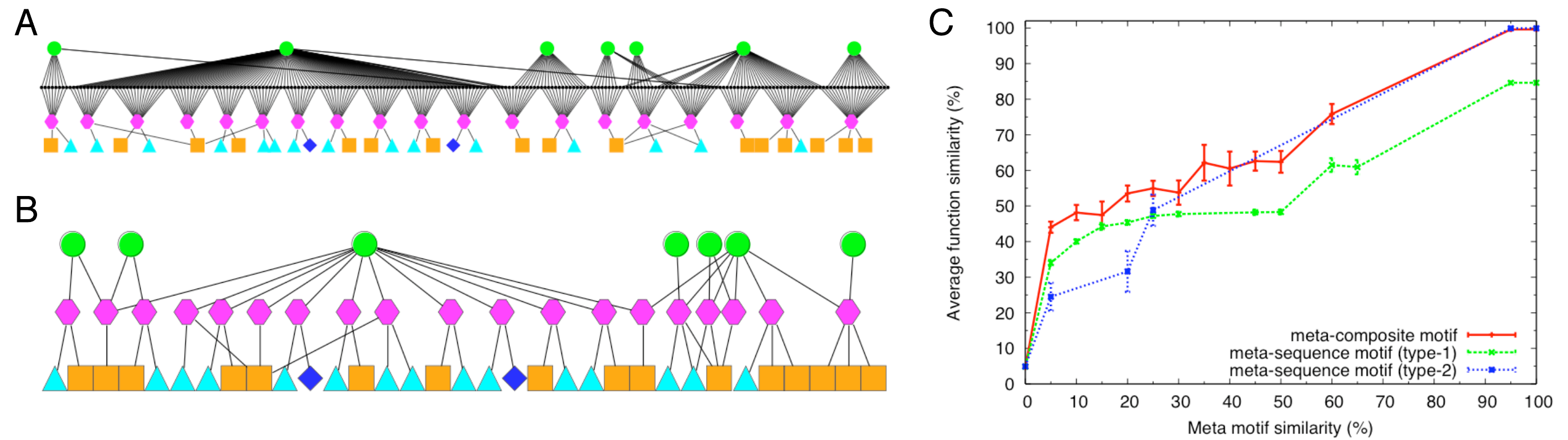}    
  \end{center}

  \caption{\label{fig:metacomp}
\textbf{Meta-composite motifs.}
A: A meta-composite motif is defined as a set of all composite motifs (hexagons in magenta) associated with particular UniProt functions (green circles). 
The associations are defined through individual protein subunits (black dots); see text for the detailed definitions.
Each composite motifs are associated with elementary motifs for non-polymer (triangles in cyan), protein (rectangles in orange), or nucleic acid (diamonds in blue) binding sites (c.f. Fig. \ref{fig:motif}). 
B: A simplified representation of the diagram shown in A.
C: Average function similarity as a function of meta-composite motif similarity
or meta-sequence motif (type-1 and type-2) similarity. 
}
\end{figure}

While each composite motif describes a particular state of a protein
subunit, any biological process is realized as a series of interaction
patterns.  In this sense, composite motifs only represent snapshots of
biological processes. To have a more integrative view of biological
processes, we define \emph{meta-composite motifs} by grouping all the
composite motifs associated with particular functions 
(Fig. \ref{fig:metacomp}A,B). 
For 3,359 UniProt functions, 2,760 meta-composite motifs were identified. The
number of composite motifs associated with meta-composite motifs
ranged from 1 to 157, with the average of 2.39 (S.D 4.62).
While the same UniProt function implies the same meta-composite motif 
by definition, the converse does not hold in general as there are more 
functions than meta-composite motifs. 
Meta-composite motifs thus allow us to understand protein functions as 
an ensemble of snapshots of ligand-bound states of proteins. 
For comparison, we analogously defined meta-sequence motifs by associating each 
function with corresponding sequence clusters (complete linkage).
We defined two types of sequence clusters, the one (type-1 sequence 
cluster) is based solely on BLAST E-value cutoff of 0.05, the other 
(type-2 sequence cluster) is based on sequence identity 
cutoff of 100\%. Thus, the former sequence clusters include a wide 
range of homologous sequences while the latter include only (almost) 
identical sequences. 

We then compared the meta-composite motif or meta-sequence motif
similarities with function similarity (Fig. \ref{fig:metacomp}C).  It
is not surprising that the function similarity appears lower for the
meta-composite motif similarity than for composite motif similarity
because, by definition, different meta-composite motifs always have
different functions while different composite motifs may have
identical functions.  Although the differences are small, we can still
observe that similar meta-composite motifs imply more similarity in
functions than either type-1 or type-2 meta-sequence motifs (Fig. \ref{fig:metacomp}C).

It is also noted that the average size of meta-composite motifs
(2.39$\pm$4.62) is statistically significantly greater than those of
meta-sequence motifs (1.88$\pm$4.42 for type-1, 1.86$\pm$3.43 for
type-2). This indicates that the composite motifs more finely dissect
protein functions than the sequence clusters.

\subsection*{Network structure of meta-composite motifs in biological processes}

\begin{figure}[!ht]
  \begin{center}
    \includegraphics[width=.8\textwidth]{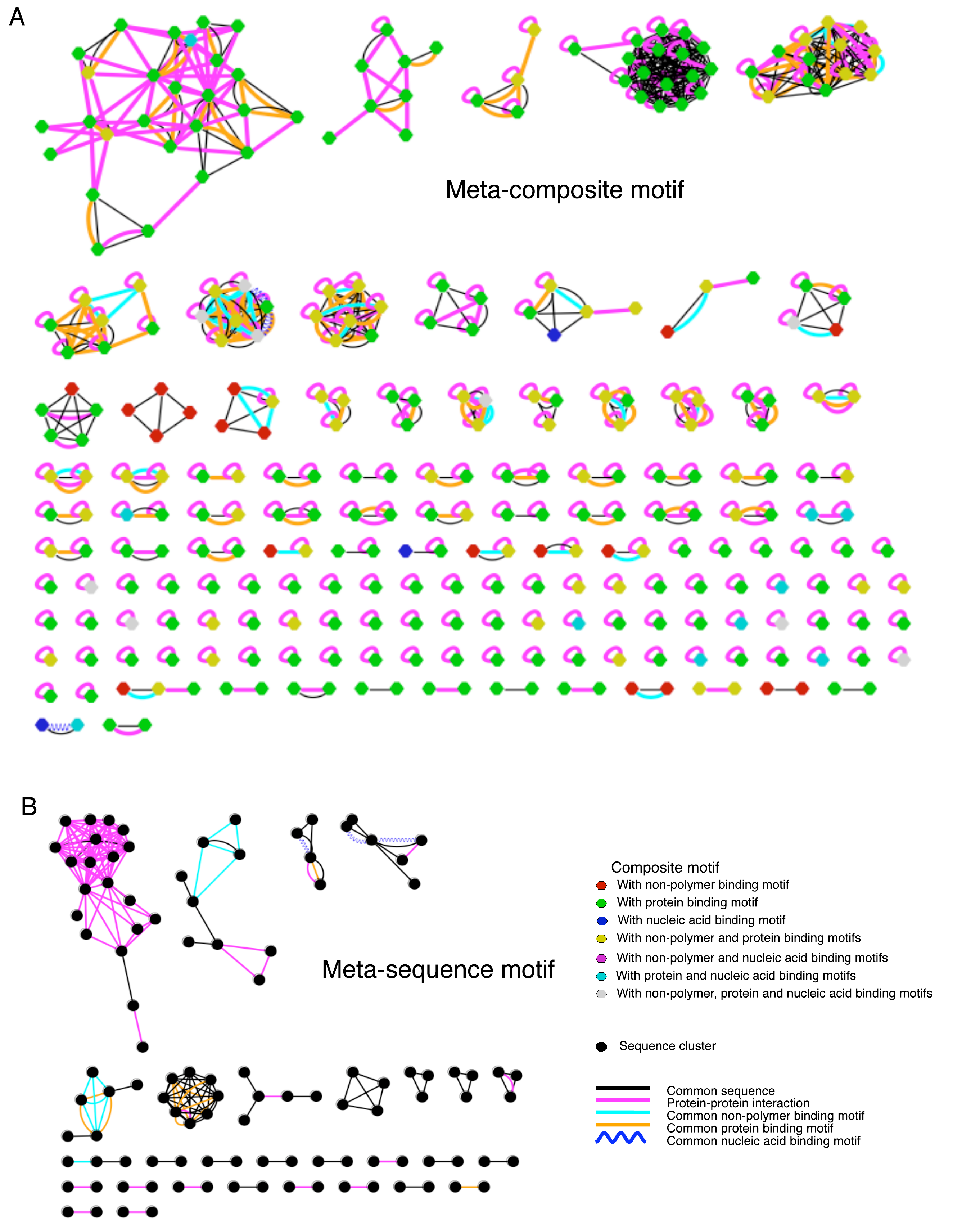}    
  \end{center}
  \caption{\label{fig:eg}
\textbf{Network structure of the meta motif for 
biological process.}  Examples of a meta-composite motif (A)
and a type-1 meta-sequence motif (B) for the UniProt biological
process ``Transcription.''  A: The meta-composite motif, i.e., the
set of composite motifs (colored hexagons) associated with
Transcription. B: type-1 meta-sequence motif, i.e., the set of
type-1 sequence clusters associated with the same keyword.  
}
\end{figure}

Since the meta-composite motifs are defined by grouping together all
composite motifs associated with particular functions, they are more
suitable for analyzing, rather than predicting, protein functions in
terms of interaction states of proteins. For example, we can identify
a meta-composite motif for the UniProt keyword ``Transcription''
(Fig. \ref{fig:eg}A), and subsequently connect the constituent
composite motifs (nodes) based on relations such as common elementary
motifs or common sequences. When a protein in one composite motif
interacts with another protein in another (possibly the same)
composite motif, an edge representing protein-protein interaction can
be also drawn.  In the case of composite motifs, nodes may be also
characterized according to their constituent elementary motifs (i.e.,
interaction states).  We can observe a variety of interaction states
of nodes and relations between nodes. If two nodes share
identical sequences, it reflects a transition between different
interaction states, possibly changing their atomic structures. 
For example, there are PDB entries of human cellular 
tumor antigen p53 with or without bound DNA (e.g., PDB 1UOL \cite{1UOL} 
and 2AC0 \cite{2AC0}) which share the same elementary motif for zinc binding
but have different composite motifs depending on the presence or
absence of the elementary motif for DNA binding. Similarly, there are
PDB entries of yeast RNA polymerase II with or without bound DNA/RNA in
which the subunit RPB2 (e.g., PDB 1I3Q \cite{1I3Q}, chain B and 1Y1W
\cite{1Y1W}, chain B) share some elementary motifs for protein
binding, but other corresponding protein binding sites belong to
different elementary motifs due to slight conformational changes (not shown), 
and an elementary motif for binding DNA is present in only one of
the entries; thus these subunits identical in amino acid sequence 
have different composite motifs which are connected by edges of the 
common protein binding motifs and of the common sequence. 
Such description is not possible with meta-sequence motifs
(Fig. \ref{fig:eg}B) because sequence similarity alone cannot
discriminate different interaction states.

\begin{figure}[!ht]
  \begin{center}
    \includegraphics[width=.4\textwidth]{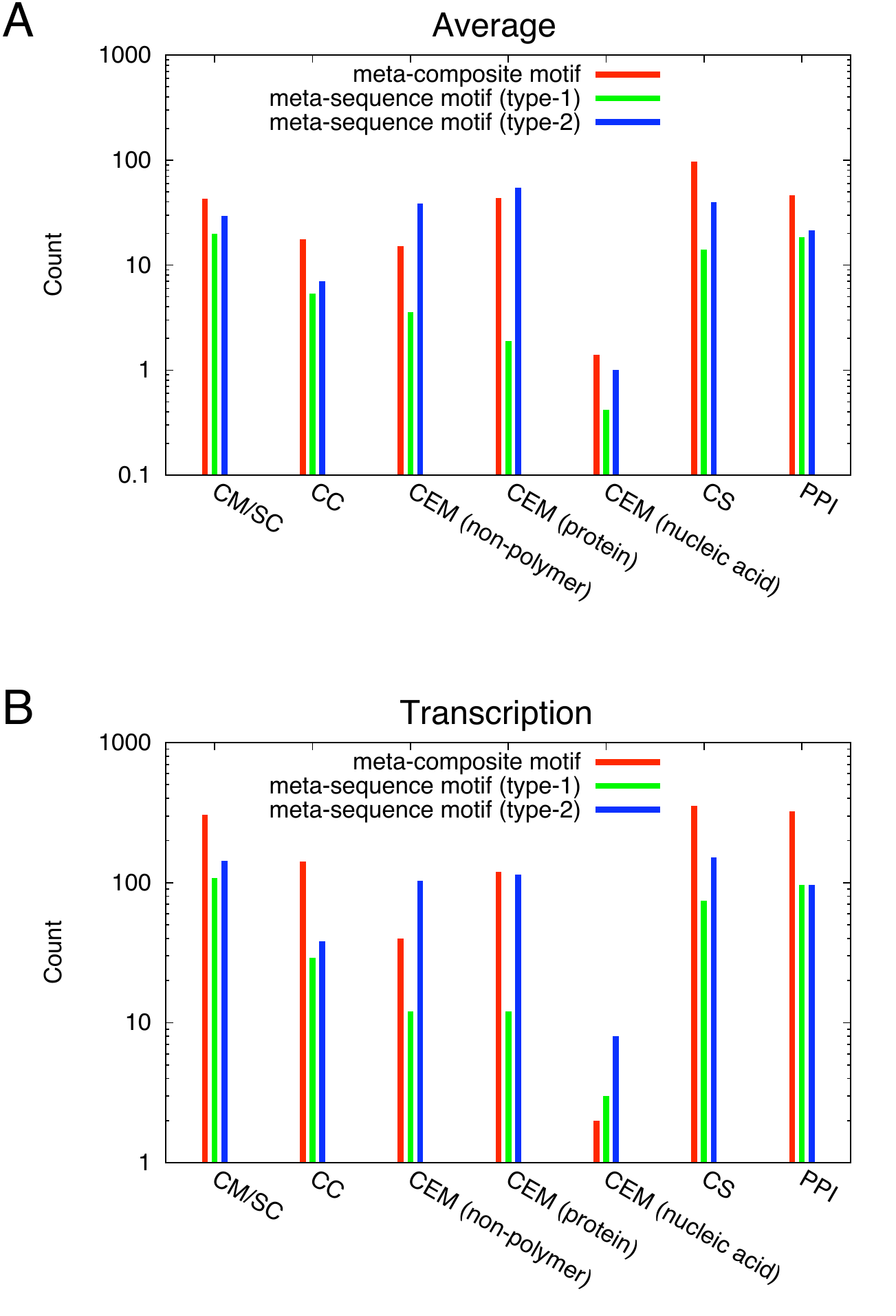}    
  \end{center}
  \caption{\label{fig:metahisto}
\textbf{Characteristics of meta motif networks.} A: Average counts of 
composite motifs or sequence clusters (denoted CM/SC), 
connected components (CC) as well as edges representing sharing of common elementary motifs (CEM) for
non-polymer, protein and nucleic acid binding sites, common sequences (CS) and
protein-protein interactions (PPI). B: The same counts for nodes and
various edges, but only for the meta motifs for the UniProt keyword
``Transcription'' (corresponding to the diagrams in Fig. \ref{fig:eg}).}
\end{figure}

To evaluate the properties of networks of meta motifs more generally
and more quantitatively, we identified the meta motif for each
upper-most keyword in the hierarchy of the UniProt Biological process
category, and compared various network characteristics of
meta-composite motifs against those of meta-sequence motifs
(Fig. \ref{fig:metahisto}).  On average (Fig. \ref{fig:metahisto}A),
meta-composite motifs include more nodes (i.e., composite motifs),
more connected components, as well as more connections between nodes
representing common sequences (identified by the UniProt
accession) and protein-protein interactions, compared to both type-1
and type-2 meta-sequence motifs.  In particular, the increased number
of edges representing common sequences indicates that many identical
proteins are split into different composite motifs. The same trend is
also observed for a particular meta-composite motif obtained for the
keyword ``Transcription'' (Fig. \ref{fig:metahisto}B).  As expected,
the type-1 meta-sequence motifs exhibit rather poor characteristics in
most aspects because many homologs are grouped into large clusters so
that differences in interaction states of proteins cannot be
differentiated. While the type-2 meta-sequence motifs sometimes
contain more edges for common elementary motifs, this is simply
because many elementary motifs shared among homologous proteins are
split into different sequence clusters irrespective of interaction
states, which is reflected in the lower number of edges representing
common sequences. Thus, the classification of proteins in terms of
composite motifs allows us to inspect the organization of proteins
involved in individual biological processes.

In summary, the observation that meta-composite motifs have more counts in
nodes, connected components, common sequences and protein-protein
interactions implies that meta-composite motifs discriminate the
subtle differences in the interaction states or conformations of the
proteins involved in the biological processes and such discrimination is 
not possible with meta-sequence motifs.

\section*{Discussion}
\label{sec:discussion}
Structural classifications of proteins have been traditionally
targeted at elucidating the universality of protein architectures
based on the notion of structural domains. As such, it is not
necessarily suitable for analyzing specific functions of particular proteins
\cite{PetreyANDHonig2009}. In other words, the current protein
structure classifications, for a good reason, ignore the differences
among protein structures within the same families or folds. The examples 
shown in Figs. \ref{fig:motif} and \ref{fig:examples} clearly show that 
although those proteins share the same folds, they have varied functions. 
Such limitations of fold classifications with respect to specific assignment 
of protein functions have been known for some time \cite{NaganoETAL2002}.  
Recently, seemingly minute differences within protein folds are being
recognized as determinants of functional specificity as exemplified by
the concept of ``embellishments'' proposed by Orengo and coworkers
\cite{DessaillyETAL2009, RedfernETAL2009}. Although it is often
assumed that domains are the units of functions, there are inherent
limitations in this assumption.  For example, it has been known that
the combination of domains generates new functions
\cite{BashtonANDChothia2007}, therefore it is questionable to assign
one function to one domain. Furthermore, the very definition of
domains is problematic as there exists no universally accepted
definition of domains \cite{HollandETAL2006}.

In this study, we avoided the complications regarding the definition
of domains, and directly analyzed the atomic structures of
binding sites irrespective of overall topology or homology of
proteins. Nevertheless, it has been previously shown that thus
identified elementary structural motifs are mostly confined within
homologous protein
families \cite{KinjoANDNakamura2009,KinoshitaETAL1999}, especially
for protein binding sites \cite{KinjoANDNakamura2010}. 
In this sense, the classification
of binding site structures are effectively not very different from the
traditional protein classifications.  However, by combining the
elementary motifs found in individual subunit structures solved under
different experimental conditions, it becomes possible to specify a
particular interaction state for a particular subunit. Thus, the
classification of proteins based on composite motifs differs from the
traditional classification schemes in that the notion of the
composite motif allows us to explicate the universality of binding site
structures and the diversity of their combinations at the same
time. It should be stressed that the redundancy of the current PDB is
essential for identifying elementary and composite motifs
since the diversity of atomic structures is not negligible even for
highly homologous or identical
proteins \cite{KinjoANDNakamura2009,ThanguduETAL2010}. In addition,
different interaction states of the same protein are also useful for
characterizing conformational transitions \cite{KoikeETAL2008,AmemiyaETAL2011,OkazakiANDTakada2011}.

We have demonstrated that the similarity between composite motifs of
proteins well indicates the similarity between their functions
(Figs. \ref{fig:pred}A,B).  A recent study also indicates that the
integration of non-polymer and protein binding sites enhances the
detection of functional specificity \cite{RausellETAL2010}.  These
results manifest the importance of the context-dependent combination
of ligand binding motifs for understanding protein functions.  The
application of composite motifs to function prediction, however,
requires some caveats. In case when we know a protein structure with
bound ligands, we first need to identify the elementary motifs to
which the binding sites belong to. But it may not be always possible to
identify all the necessary elementary motifs.  In case when we only
have a protein structure in its ligand-free form, it is necessary to
predict its binding sites if any should exist. In this case, we need
to rely on prediction based on prediction, which necessarily leads to
low accuracy. While this limitation is inherent in any annotation
transfer approaches, it is more stringent on the one based on
composite motifs because it requires more interaction states to be
solved for similar proteins.  In any case, it is preferable to
accumulate more structures in the PDB, not only those of completely
novel folds, but also those of known folds but in new ligand-bound
forms. It is worth noting that the function prediction by composite motif
similarity is not based on supervised learning or parameter fitting so that 
the results obtained here should hold mostly valid for newly 
solved structures to the extent that the distribution of functionally 
characterized proteins in the PDB stays the same.

By grouping the composite motifs associated with particular functions,
we defined meta-composite motifs. It was demonstrated that the
description based on meta-composite motifs provided us with a detailed
annotation of biological processes
(Figs. \ref{fig:metacomp},\ref{fig:eg}).  By describing biological
processes in terms of composite motifs rather than individual
structures, we can abstract the pattern of interactions so that the
commonality and specificity of the interactions in different contexts,
such as species or pathways, for example, can be delineated.
Although there are currently some limitations in this description,
such as the absence of temporal relation between composite motifs or
the lack of experimental structures for some possible transient
complexes, these limitations may be overcome in the future by
complementing meta-composite motifs with other experimental
information such as gene/protein expression or interactome analyses.

In summary, we have introduced composite motifs that well describe
protein functions based on the context-dependent combinations of 
structural patterns of binding sites, and provide a useful means to 
describe the atomic details of biological processes.

% You may title this section "Methods" or "Models". 
% "Models" is not a valid title for PLoS ONE authors. However, PLoS ONE
% authors may use "Analysis" 
\section*{Materials and Methods}
\subsection*{Data set}
We have used all the PDB entries as of December 29, 2010 (70,231
entries). All the biological units were generated for each entry as
annotated in the PDBML files \cite{PDBML}, except for those with
icosahedral, helical, or point symmetries (mostly viruses). 
For the latter, only the corresponding (icosahedral, etc.) asymmetric 
units were used. Entries without annotated biological units were treated 
as they are given. Some PDB entries contain more than one biological unit
all of which were used in the present study since alternative oligomeric states
may (or may not) be biologically relevant. The biological units 
in the PDB are defined by authors and/or software 
(PQS\cite{PQS} and/or PISA\cite{PISA}).
In total, 197,690 subunits in 79,826 biological units contained 
at least one ligand binding site.

A ligand binding site of a subunit is defined as a set of at least 10 
atoms in the subunit that are in contact with some atoms of a ligand within 
5\AA{} radius. In this study, ligands include non-polymers, proteins, and 
nucleic acids. The non-polymer ligands are those annotated as such in the 
PDBML \cite{PDBML} files, but water molecules were discarded.
The protein ligands are those annotated as ``polypeptide(L)'' with at least 
25 amino acid residues. 
The nucleic acid ligands are those annotated as 
``polydeoxyribonucleotide,'' ``polyribonucleotide'' or
``polydeoxyribonucleotide/polyribonucleotide hybrid.''

\subsection*{Similarity between binding site structures}
To compare binding site structures, we used the GIRAF structural search
and alignment program \cite{KinjoANDNakamura2007} with some modifications
to enable faster database search and flexible alignments (unpublished). 
GIRAF produces an
atom-wise alignment for a pair of binding sites. After all-against-all
comparisons of binding sites, elementary motifs were defined as
complete-linkage clusters with a cutoff GIRAF score
\cite{KinjoANDNakamura2007} of 15, as in our previous studies
\cite{KinjoANDNakamura2009,KinjoANDNakamura2010}. The cutoff value was chosen 
so that the largest cluster did not predominate all the other clusters due to 
the ``phase transition'' of the similarity networks \cite{Bollobas, KinjoANDNakamura2009}.
The GIRAF score is defined as 
\begin{equation}
  \label{eq:girafscore}
  G(A,B) = \frac{N_{A,B}\sum_{a}w(\mathbf{x}^A_a,\mathbf{x}^B_a)}{\min\left[N_A,N_B\right]}
\end{equation}
where $N_A$ and $N_B$ are the number of atoms of the binding sites $A$
and $B$ respectively, and $N_{A,B}$ is the number of aligned atom
pairs.  The weight $w(\mathbf{x}^A_a , \mathbf{x}^{B}_a)$ for the
aligned atom pairs $\mathbf{x}^A_a$ and $\mathbf{x}^{B}_a$ ($a =
1, \cdots, N_{A,B}$) is defined as
\begin{equation}
  \label{eq:wpair}
  w(\mathbf{x}^A_a,\mathbf{x}^B_a) = \max\left[1 - d(\mathbf{x}^A_a,\mathbf{x}^B_a)/d_c, 0\right]
\end{equation}
where $d(\mathbf{x}^A_a,\mathbf{x}^B_a)$ is the distance between two
atoms in a superimposed coordinate system and the cutoff distance
$d_c$ is set to 2.5 \AA{}.

Clusters with less than 10 members were excluded in this study because
structural similarity in small clusters may be coincidental.
In fact, when there were
protein pairs not detected by BLAST within a cluster, the fraction of
such pairs was 79\% on average for clusters with less than 10 members
while that for clusters with at least 10 members was 36\%. Although motifs 
shared between remote homologs or non-homologs may provide interesting
examples, we expect many of them are not biologically relevant.

The raw GIRAF score largely depends on the size of binding sites. 
Therefore, when
comparing binding site similarity with function similarity, we used a
normalized similarity measure so that binding sites of varying sizes can
be compared on the same scale.  Let $N_A$, $N_B$ and $N_{A,B}$ be 
defined as above, then the normalized similarity $S(A,B)$
between the binding sites $A$ and $B$ is defined as
\begin{equation}
  \label{eq:simif} S(A,B) = 100 \times \frac{2 N_{A,B}}{N_A + N_B}~~(\%).
\end{equation}

\subsection*{Functions defined by UniProt keywords}
For each subunit in the data set, the corresponding UniProt
\cite{UniProt} accession identifier was obtained from the struct\_ref
category of the PDBML file. In total, 186,791 subunits with at least 1 
ligand binding site in the PDB were
annotated by UniProt. For thus identified UniProt entries, their
keywords were extracted.  The UniProt keywords are a set of controlled
vocabulary to describe the properties of proteins and they are
organized in a hierarchical order. In most cases, these keywords 
are manually assigned by curators, hence they are expected to be 
more reliable. This is in contrast to the Gene Ontology 
annotations (http://geneontology.org) for the PDB which are 
mostly automatically annotated and are likely to contain 
a large number of erroneous annotations. 

For each subunit, all the keywords
annotated either explicitly or implicitly via the keyword hierarchy,
were extracted except for those belonging to the Technical term,
Disease, or Domain categories.  We define the function of a subunit as
the set of the UniProt keywords associated with it. In other words,
two subunits whose associated sets of keywords are exactly identical
are defined to have the same function.  In total, 7,991 UniProt
functions were defined.  The similarity between two UniProt functions
are defined by the Jaccard index between the sets of keywords
associated with the functions (see below, Eq. \ref{eq:jac})

\subsection*{Similarity between two sets}
The similarity measures for composite motifs, functions or meta motifs are 
based on comparison between two sets. Given the sets $A$ and $B$, their 
similarity is defined by the Jaccard index $J(A,B)$:
\begin{equation}
  \label{eq:jac}
  J(A,B) = 100\times \frac{|A \cap B|}{|A \cup B|}~~(\%).
\end{equation}
For a given composite motif, function, meta-composite motif 
or meta-sequence motif, the set consists of elementary motifs, 
UniProt keywords, composite motifs, or sequence clusters, respectively.

\subsection*{Sequence clusters}
To define meta-sequence motifs, complete-linkage clustering was applied to
the result of an all-against-all BLAST \cite{AltschulETAL1997} comparison 
with two different criteria. In one case, all pairs of sequences in a 
cluster must have BLAST E-value of at most 0.05. This resulted in 3,327 
clusters with at least 10 members. These clusters are referred to as type-1 
sequence clusters. In the other case, all pairs of sequences 
in a cluster must have 100\% sequence identity as well as E-value of at 
most 0.05. This resulted in 4,594 clusters with at least 10 members,
which are referred to as type-2 sequence clusters.
When BLAST produces more than one alignment for a pair of sequences, 
the alignments were integrated into one alignment as long as they were 
mutually consistent.

\subsection*{Comparison between motif similarity and function similarity}
Although we did not use any representative set for defining elementary
and composite motifs based on sequence similarity, we did use
representatives of motifs and sequences when their similarities were
compared with function similarity (c.f., Figs. \ref{fig:pred} and
\ref{fig:metacomp}C) in order to reduce the bias due to different
sizes of clusters. For composite motifs, a representative was
randomly selected from each composite motif. For binding sites, 
a representative was randomly selected from each elementary
motif.  For protein sequences, a representative was randomly selected
from each type-2 sequence cluster.  Average function similarities for
a given range of motif, binding site or sequence similarity
(Fig. \ref{fig:pred}) were calculated for 10 sets of randomly
selected representatives and the standard deviations of the average
function similarity are shown as error bars. Only those points with 
at least 500 (50 for nucleic acid binding sites) samples on 
average are shown in Figs. \ref{fig:pred}A,B.

For meta-composite and meta-sequence motifs, 50 \% of the all observed
pairs of meta motifs were randomly selected and the average function
similarity was calculated. This procedure was iterated 10 times, and
data points with at least 10 samples are reported with the standard
deviation of the average values in Fig. \ref{fig:metacomp}C.

We have confirmed that selecting different random sets of representatives in 
all the above cases did not alter the results significantly.

\subsection*{Downloadable data}
The results of all-against-all comparison of binding sites and
classifications are made available for download at
http://pdbj.org/giraf/cmotif/.

% Do NOT remove this, even if you are not including acknowledgments
\section*{Acknowledgments}
We thank Toshiaki Katayama and Hideki Hatanaka for helping with
uploading data to TogoDB.  

%\section*{References}
% The bibtex filename
%\bibliography{refs,mypaper}

\section*{Figure Legends}

%\section*{Tables}
%\begin{table}[!ht]
%\caption{
%\bf{Table title}}
%\begin{tabular}{|c|c|c|}
%table information
%\end{tabular}
%\begin{flushleft}Table caption
%\end{flushleft}
%\label{tab:label}
% \end{table}

\end{document}